\documentclass[%
 reprint,
 amsmath,amssymb,
 aps,
]{revtex4-1}

\usepackage{graphicx}% Include figure files
\usepackage{dcolumn}% Align table columns on decimal point
\usepackage{bm}% bold math
\begin{document}

\preprint{APS/123-QED}

\title{Direct numerical simulation of dispersed particles in a compressible fluid}% Force line breaks with \\

\author{Rei Tatsumi}
 \email{tatsumi@cheme.kyoto-u.ac.jp}
\author{Ryoichi Yamamoto}%
 \email{ryoichi@cheme.kyoto-u.ac.jp}
\affiliation{%
 Department of Chemical Engineering, Kyoto University, Kyoto 615-8510, Japan
}%

\date{\today}% It is always \today, today,

\begin{abstract}
We present a direct numerical simulation method for investigating the dynamics of dispersed particles in a compressible solvent fluid.
The validity of the simulation is examined by calculating the velocity relaxation of an impulsively forced spherical particle with a known analytical solution.
The simulation also gives information about the fluid motion, which provides some insight into the particle motion.
Fluctuations are also introduced by random stress, and the validity of this case is examined by comparing the calculation results with the fluctuation-dissipation theorem.
\end{abstract}

\pacs{Valid PACS appear here}% PACS, the Physics and Astronomy
\maketitle

%\tableofcontents

\section{\label{sec:level1}Introduction}

Particle dispersions have various unique properties, 
and an understanding of these properties is important in many fields of science and engineering.
These properties originate from the dynamics of particles, 
which are extremely complicated because of the hydrodynamic interactions among particles mediated by the motion of the surrounding fluid.
Therefore, several numerical approaches have been formulated to investigate the dynamics of such dispersions. 
As one of these approaches, a direct numerical simulation has been developed, 
wherein the hydrodynamic interactions are directly computed by simultaneously solving for the motion of the fluid and the motion of the particle. 
In recent years, we have developed an efficient direct numerical simulation scheme for dispersions, which is called the smoothed profile method (SPM)~\cite{B1,B2}.
In this scheme, the sharp interface between the fluid and the particles is replaced by a smoothed interface using a continuous profile function.
The Navier-Stokes equations are solved for the fluid motion on a fixed square grid, 
and Newton's and Euler's equations of motion for the particles are solved simultaneously while considering the momentum exchange between the fluid and the particles.

The hydrodynamic interactions are transmitted in two ways: via viscous
momentum diffusion and via sound propagation.
The time scale of viscous diffusion over the particle size 
is $\tau_\nu = a^2/\nu$, 
and that of sound propagation is $\tau_c = a/c$, 
where $a$ is the particle radius, $\nu$ is the kinematic viscosity, and $c$ is the speed of sound in the fluid.
Here, we will define the compressibility factor as the ratio of the two time scales:
\begin{eqnarray}
\varepsilon &=& \frac{\tau_c}{\tau_\nu} = \frac{\nu}{ac}
\label{e1-1}.
\end{eqnarray}
This factor represents the degree of influence of compressibility in the dynamics of the dispersed particles.
According to Eq.~(\ref{e1-1}), the compressibility becomes increasingly important as the particle size decreases.
Indeed, the compressibility of a system has been considered an important factor 
in molecular scale dynamics of monatomic liquids studied using hydrodynamic theories~\cite{B3,B4}.
On the other hand, in studies of particle dispersions, compressibility has rarely been considered, and an incompressible host fluid is generally assumed.
In the case of a dispersed particle of radius $a = 100\:{\rm nm}$ in water, for instance,
the compressibility factor is evaluated from $\nu = 1.0 \times 10^{-6}\:{\rm m^2/s}$ and $c = 1.5 \times 10^3\:{\rm m/s}$ to be $\varepsilon = 6.7 \times 10^{-3}$.
In many cases, researchers are interested in phenomena progressing over the time scale of viscous diffusion or even longer time scales, 
such as those relating to shear properties, electrophoresis, and sedimentation.
Therefore, the assumption of an incompressible host fluid is valid,
and most direct numerical simulation methods, including SPM, have been developed on the premise of incompressible fluids.
However, when we investigate phenomena associated with sound propagation,
such as sonic agglomeration, acoustic spectroscopy, and electroacoustic measurements,
the consideration of compressibility is required.

In the present study, we extend the SPM to compressible fluids.
Some aspects of the dynamics of a single particle in a compressible fluid have been theoretically analyzed~\cite{B3, B4, B5,B6,B7},
and we compare the simulation results obtained herein with analytical solutions to determine the accuracy of the simulation.
In particular, we consider the velocity relaxation of a spherical particle after an impulsive force is added.
The numerical simulation also gives information regarding fluid motion for which the analytical solution is unknown,
and the dynamics of the particle can be investigated from the viewpoint of the fluid dynamics.
In addition, we also consider a system with thermal fluctuations by introducing random stress, and the velocity autocorrelation function is compared with the analytical solution according to the fluctuation-dissipation theorem.

\section{Simulation Method}

\subsection{\label{sec:level2}Equations}

In the SPM, the particle-fluid boundary is replaced by a continuous interface.
For this purpose, the smoothed profile function $\phi(\boldsymbol{r},t) \in [1,0]$ is introduced.
This function represents the boundary between the fluid and particle regions: $\phi = 1$ at the particle domain, and $\phi = 0$ at the fluid domain.
The two regions are smoothly connected through thin interfacial regions with a thickness characterized by $\xi$.
The mathematical expression of $\phi$ can be found in a previous paper~\cite{B1}.
The total velocity field is defined as 
\begin{eqnarray}
\boldsymbol{v} &=& (1 - \phi)\boldsymbol{v}_f + \phi \boldsymbol{v}_p
\label{e2A-1},
\end{eqnarray}
where $\boldsymbol{v}_f(\boldsymbol{r},t)$ is the fluid velocity field, and $\boldsymbol{v}_p(\boldsymbol{r},t)$ is the particle velocity field constructed from rigid motions of the particles~\cite{B1,B2}.
When a compressible host fluid is considered, the fluid mass density is altered,
and we define the mass density field as
\begin{eqnarray}
\rho_f &=& (1 - \phi) \rho
\label{e2A-2}.
\end{eqnarray}
The auxiliary mass density field $\rho(\boldsymbol{r},t)$ is defined over the entire domain.
However, the physical fluid mass density $\rho_f(\boldsymbol{r},t)$ must be zero within the particle domain, 
and this requirement is satisfied by multiplying by $(1-\phi)$.

%In the case of an incompressible fluid, the impermeability condition is satisfied by the solenoidal condition of both of the fluid and particle velocity fields.
%The condition $\nabla \cdot \boldsymbol{v}_p = 0$ is also satisfied, even though a compressible fluid is assumed.
%Here, the mass conservation in Lagrange representation is given as
%\begin{eqnarray}
%\frac{1}{\rho} \frac{D \rho}{Dt} &=& -\nabla \cdot \boldsymbol{v},\\
%&=& -(1 - \phi) \nabla \cdot \boldsymbol{v}_f - \nabla \phi \cdot (\boldsymbol{v}_p - \boldsymbol{v}_f)
%\label{e1-3}.
%\end{eqnarray}
%The second term on the right-hand side represents the density change at the interfacial region.
%When the velocity in the particle region is higher than that in the fluid region, 
% compression occurs in front of the particle and rarefaction occurs behind the particle in the interfacial region.
%This effect prevents fluid inflow into the particle or leaks from the particle to the fluid.
%Therefore, the impermeability condition is implicitly satisfied.

The equations governing the dynamics of the dispersion system are given as hydrodynamic equations with the addition of a body force.
The hydrodynamic equations consist of three conservation laws concerning mass, momentum, and energy.
The conservation equations of mass and momentum are given as
\begin{eqnarray}
\frac{\partial \rho}{\partial t} + \boldsymbol{\nabla} \cdot \boldsymbol{m} &=& 0
\label{e2A-3},
\end{eqnarray}
\begin{eqnarray}
\frac{\partial \boldsymbol{m}}{\partial t} + \boldsymbol{\nabla} \cdot (\boldsymbol{mv}) &=& \boldsymbol{\nabla} \cdot \boldsymbol{\sigma} + \rho \phi \boldsymbol{f}_p
\label{e2A-4},
\end{eqnarray}
where $\boldsymbol{m}(\boldsymbol{r},t) = \rho(\boldsymbol{r},t) \boldsymbol{v}(\boldsymbol{r},t)$ is the momentum density field.
We consider a compressible Newtonian fluid, and the stress tensor is given by
\begin{eqnarray}
\boldsymbol{\sigma} &=& -p \boldsymbol{I} + \eta [\boldsymbol{\nabla} \boldsymbol{v} + (\boldsymbol{\nabla} \boldsymbol{v})^T] + \left( \eta_v - \frac{2}{3}\eta \right)(\boldsymbol{\nabla} \cdot \boldsymbol{v}) \boldsymbol{I}
\label{e2A-5},
\end{eqnarray}
where $p(\boldsymbol{r},t)$ is the pressure, $\eta$ is the shear viscosity, and $\eta_v$ is the bulk viscosity.
A body force $\rho \phi \boldsymbol{f}_p$ is added so that the rigidity of the particles is satisfied.
Additionally, we assume a barotropic fluid described by $p = p(\rho)$,
and the pressure gradient is proportional to that of the mass density:
\begin{eqnarray}
\boldsymbol{\nabla} p &=& c^2 \boldsymbol{\nabla} \rho
\label{e2A-6}.
\end{eqnarray}
Equations~(\ref{e2A-3})-(\ref{e2A-6}) are closed for the variables $\rho$, $\boldsymbol{m}$, and $p$;
therefore, energy conservation does not need to be considered.

The motion of the dispersed particles is governed by Newton's and Euler's equations of motion:
\begin{eqnarray}
M_i \frac{\mathrm{d}}{\mathrm{d} t} \boldsymbol{V}_i &=& \boldsymbol{F}^H_i + \boldsymbol{F}^C_i,\ \ \ \ \frac{\mathrm{d}}{\mathrm{d} t} \boldsymbol{R}_i = \boldsymbol{V}_i
\label{e2A-7},
\end{eqnarray}
\begin{eqnarray}
\boldsymbol{I}_i \cdot \frac{\mathrm{d}}{\mathrm{d} t} \boldsymbol{\Omega}_i &=& \boldsymbol{N}^H_i
\label{e2A-8},
\end{eqnarray}
where $\boldsymbol{R}_i$, $\boldsymbol{V}_i$, and $\boldsymbol{\Omega}_i$ are the position, translational velocity, and rotational velocity of the $i\mathchar`-$th particle, respectively.
The particle has a mass $M_i$ and a moment of inertia $\boldsymbol{I}_i$.
The hydrodynamic force $\boldsymbol{F}^H_i$ and torque $\boldsymbol{N}_i^H$ are exerted on the particle by the fluid,
and the force $\boldsymbol{F}^C_i$ is exerted through direct interactions among the particles.

The effect of thermal fluctuations on the dynamics of particles is important when the particle size is on the order of a micrometer or smaller.
Fluctuations are introduced through a random stress tensor $\boldsymbol{s}$, which is added in Eq.~(\ref{e2A-5}).
The random stress is a stochastic variable satisfying the fluctuation-dissipation relation~\cite{B8}:
\begin{eqnarray}
\langle s_{ij}(\boldsymbol{r}, t) s_{kl}(\boldsymbol{r}^{\prime}, t^{\prime}) \rangle &&= \nonumber \\
2 k_B T &&\eta_{ijkl} \delta(\boldsymbol{r}^{\prime} - \boldsymbol{r}) \delta(t^{\prime} - t)
\label{e2A-9},
\end{eqnarray}
where $k_B$ is the Boltzmann constant, $T$ is the temperature, and
\begin{eqnarray}
\eta_{ijkl} &=& \eta(\delta_{ik}\delta_{jl} + \delta_{il}\delta_{jk}) + \left( \eta_v - \frac{2}{3} \eta \right)\delta_{ij}\delta_{kl}
\label{e2A-10}.
\end{eqnarray}
Brownian motion of the dispersed particles is induced by random stress acting on the fluid.
Thermal fluctuations can be introduced using the Langevin approach, where random forces are exerted on the particles~\cite{B9}.
However, this approach does not accurately represent the short-time dynamics of the system because the time autocorrelation of the hydrodynamic force acting on the particles is neglected.
Therefore, the fluctuating hydrodynamics approach is more appropriate for investigating dynamics over the time scale of sound propagation.

\subsection{Simulation procedure}

Here, the time-discretized evolution of the equations is derived.
The time step $t_n$ represents the $n\mathchar`-$th discretized time, 
and the time step change from $t_n$ to $t_{n+1} = t_n+h$ will be considered.
The time evolution of the system is determined through the following steps:
\begin{itemize}
\item[(i)]
The mass and momentum density changes associated with sound propagation are calculated as
\begin{eqnarray}
\rho^{n+1} &=& \rho^n - \int^{t_n+h}_{t_n} \mathrm{d}t \boldsymbol{\nabla} \cdot \boldsymbol{m}, \\
\boldsymbol{m}^{\ast} &=& \boldsymbol{m}^n - c^2 \int^{t_n+h}_{t_n} \mathrm{d}t \boldsymbol{\nabla} \rho
\label{e2B-1}.
\end{eqnarray}
When we assume a periodic boundary condition and use the Fourier spectral method,
a semi-implicit scheme becomes feasible~\cite{B10}.
This situation eases the restriction on the time increment for a small compressibility factor $\varepsilon$.
\item[(ii)]
The time evolution of the advection and viscous diffusion terms are calculated as
\begin{eqnarray}
\boldsymbol{m}^{\ast \ast} &=& \boldsymbol{m}^{\ast} + \int^{t_n+h}_{t_n} \mathrm{d}t \boldsymbol{\nabla} \cdot (\boldsymbol{\tau} - \boldsymbol{mv})
\label{e2B-2},
\end{eqnarray}
where $\boldsymbol{\tau}$ is the dissipative stress given by $\boldsymbol{\sigma} = -p \boldsymbol{I} + \boldsymbol{\tau}$.
\item[(iii)]
In concert with the advection of the particle domain,
the position of each dispersed particle evolves as
\begin{eqnarray}
\boldsymbol{R}_i^{n+1} &=& \boldsymbol{R}^n_i + \int^{t_n+h}_{t_n} \mathrm{d}t \boldsymbol{V}_i
\label{e2B-3}.
\end{eqnarray}
\item[(iv)]
The hydrodynamic force and torque are derived by considering the conservation of momentum.
The time-integrated hydrodynamic force and torque are computed as
\begin{eqnarray}
\int^{t_n+h}_{t_n} \mathrm{d}t \boldsymbol{F}^H_i &=& \int \mathrm{d} \boldsymbol{r} \phi^{n+1}_i (\boldsymbol{m}^{\ast \ast} - \rho^{n+1} \boldsymbol{v}^n_p)
\label{e2B-4},
\end{eqnarray}
\begin{eqnarray}
\int^{t_n+h}_{t_n} \mathrm{d}t \boldsymbol{N}^H_i &&= \nonumber \\
\int \mathrm{d} \boldsymbol{r} [(\boldsymbol{r} - &&\boldsymbol{R}^{n+1}_i) \times \phi^{n+1}_i (\boldsymbol{m}^{\ast \ast} - \rho^{n+1} \boldsymbol{v}^n_p)]
\label{e2B-5}.
\end{eqnarray}
With these and other forces acting on the particles, the translational and rotational velocity of each dispersed particle evolve as
\begin{eqnarray}
\boldsymbol{V}^{n+1}_i &=& \boldsymbol{V}^n_i + M^{-1}_i \int^{t_n+h}_{t_n} \mathrm{d}t (\boldsymbol{F}^H_i + \boldsymbol{F}^C_i)
\label{e2B-6},
\end{eqnarray}
\begin{eqnarray}
\boldsymbol{\Omega}^{n+1}_i &=& \boldsymbol{\Omega}^n_i + \boldsymbol{I}^{-1}_i \cdot \int^{t_n+h}_{t_n} \mathrm{d}t \boldsymbol{N}^H_i
\label{e2B-7}.
\end{eqnarray}
\item[(v)]
The updated velocity of the particle region is imposed on the velocity field as the body force $\rho \phi \boldsymbol{f}_p$.
\begin{eqnarray}
\boldsymbol{m}^{n+1} &=& \boldsymbol{m}^{\ast \ast} + \int^{t_n+h}_{t_n} \mathrm{d}t \rho \phi \boldsymbol{f}_p,
\end{eqnarray}
\begin{eqnarray}
\int^{t_n+h}_{t_n} \mathrm{d}t \rho \phi \boldsymbol{f}_p &=& \phi^{n+1} (\rho^{n+1}\boldsymbol{v}^{n+1}_p - \boldsymbol{m}^{\ast \ast})
\label{e2B-8}.
\end{eqnarray}
\end{itemize}

In the case of an incompressible fluid, the pressure is spontaneously determined by the solenoidal condition of the velocity field.
On the other hand, in the present case, the pressure or mass density varies independently of the velocity field.

\section{Simulation Results}

Numerical simulations are performed for a three-dimensional box with periodic boundary conditions.
The space is divided by meshes of length $\Delta$, 
which is the unit length.
The units of the other physical quantities are defined by combining $\eta = 1$ and $\rho_0 = 1$ with $\Delta$, 
where $\rho_0$ is the fluid mass density at equilibrium.
The system size is $L_x \times L_y \times L_z = 128 \times 128 \times 128$.
The other parameters are set as $a = 4$, $\xi = 1$, $\rho_p = 1$, $\eta_v = 0$, and $h = 0.01$,
where $\rho_p$ is the particle mass density.
We performed simulations of the dynamics of an isolated spherical particle in a fluid in two situations.
First, we investigated the relaxation response of a particle with an impulsive force.
Second, we consider the velocity autocorrelation function of a particle
with thermal fluctuations.

Because the input particle radius $a=4$ is not necessarily equal to the
effective hydrodynamic radius $a^{\ast}$ of the particle
represented by the smoothed profile function $\phi$ with a fuzzy
interface of thickness $\xi$, 
we calculated $a^{\ast}$ from 
%The set particle radius has uncertainty due to the ambiguous particle-fluid boundary in the SPM.
the drag force acting on the spherical particle moving at velocity 
$V$, which is analytically given by $F_D = 6\pi\eta aK(\phi)V$, 
where $K(\phi)$ represents the effect of the periodic boundary condition depending on the volume fraction of the particle $\phi$~\cite{B11}.
The effective radius was evaluated as $a^{\ast} = 3.87$ in the 
present simulations, and therefore, the momentum diffusion time 
is estimated to be $\tau_\nu = a^{\ast 2}/\nu$ in this case.
When we compare the present simulation results with the analytical solutions, 
the corrected particle density $\rho_p^{\ast} = (a^{\ast}/a)^3 \rho_p$ and the compressibility factor $\varepsilon^{\ast} = (a/a^{\ast}) \varepsilon$ of the analytical solutions are employed.

\subsection{Relaxation}

\begin{figure}[tbp]
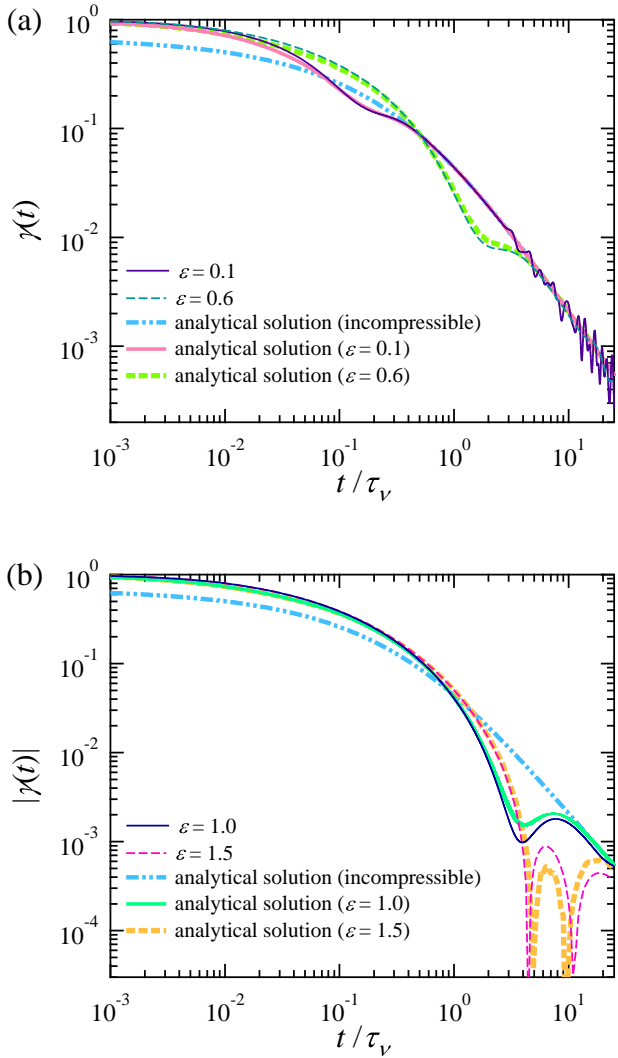

\includegraphics[width=90mm]{fig_1a.eps} \\
\includegraphics[width=90mm]{fig_1b.eps}
\caption{\label{f3} (Color online) The velocity relaxation function for various compressibility factors: (a) $\varepsilon = 0.1$, 0.6, (b) 1.0, and 1.5.
The bold lines illustrate the analytic solutions for various compressibility factors, 
and the bold dashed double-dotted line shows the analytic solution for an incompressible fluid.
}
\end{figure}

\begin{figure*}[tbp]
\includegraphics[width=40mm]{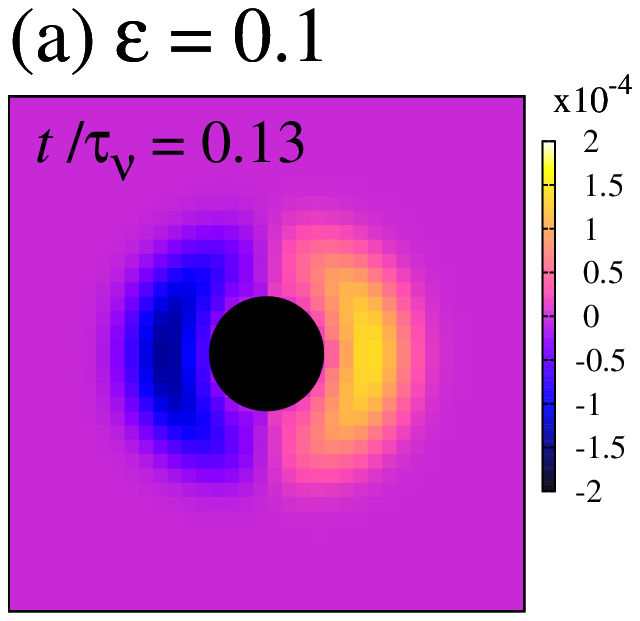}
\includegraphics[width=40mm]{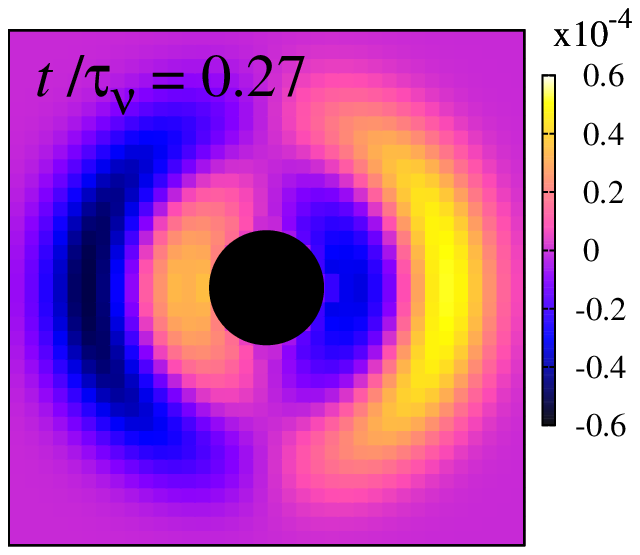}
\includegraphics[width=40mm]{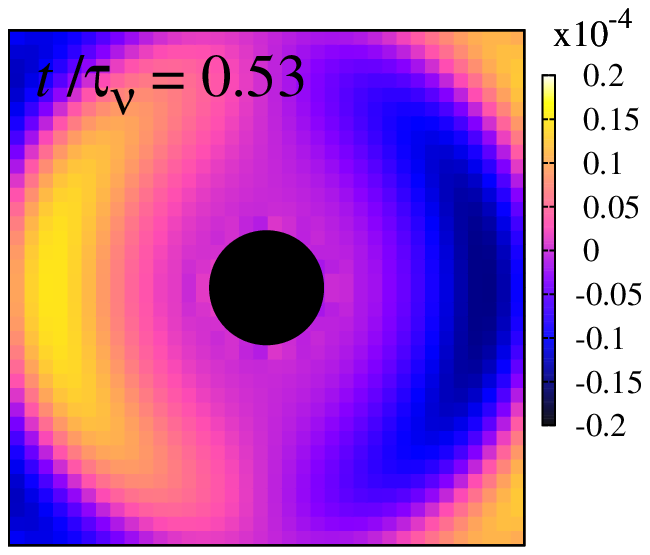} \\
\includegraphics[width=40mm]{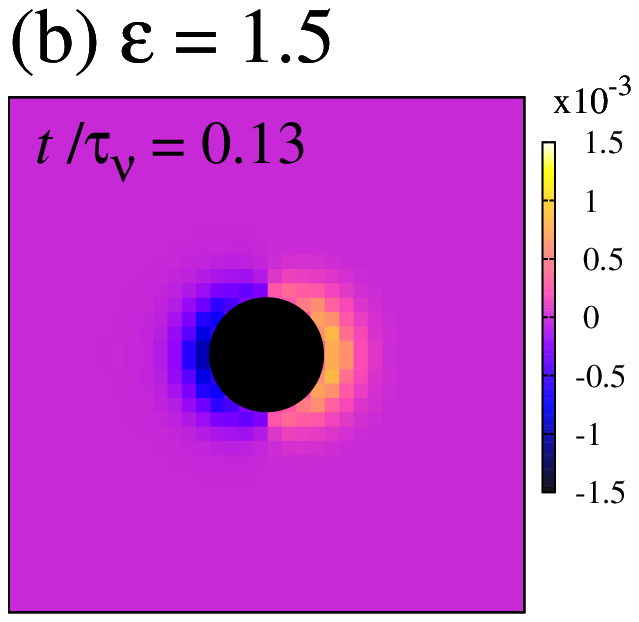}
\includegraphics[width=40mm]{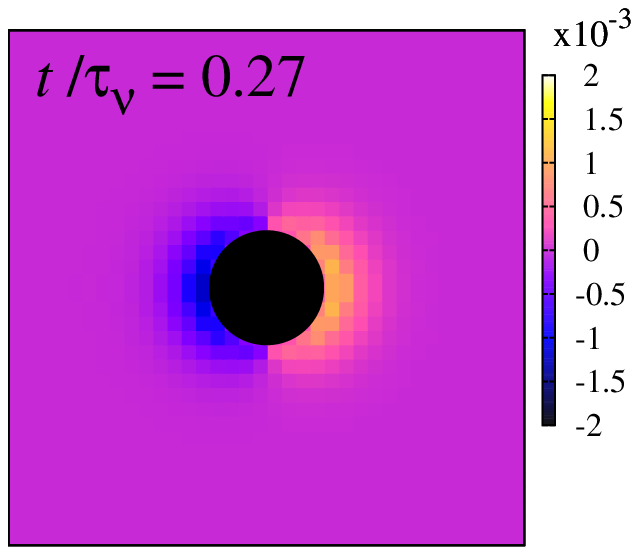}
\includegraphics[width=40mm]{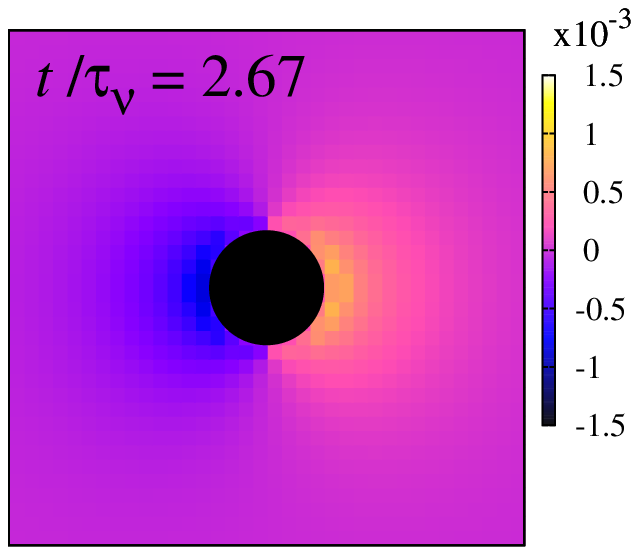}
\caption{\label{f4} (Color online) Time evolution of the fluid density deviation around the particle.
The compressibility factors are (a) $\varepsilon = 0.1$ (upper row) and (b) $\varepsilon = 1.5$ (lower row).
The color scale (gray scale) represents negative (darker) to positive (lighter) density deviation.
The black circle represents the particle.
The direction of the initial particle velocity is right in these pictures.
}
\end{figure*}

Let us consider a single spherical particle in a fluid at rest.
We will investigate the relaxation of the particle velocity after exerting an impulsive force at the center of the particle.
The impulsive force is assumed to be small; in other words, we will consider the motion of the particle and the fluid for low Reynolds and Mach numbers.
We set the impulsive force to achieve an initial particle Reynolds number of $\rm{Re}_p = 0.01$.
Here, we introduce the velocity relaxation function as the normalized velocity change of the particle as
\begin{eqnarray}
\boldsymbol{V}(t) &=& \frac{\boldsymbol{P}}{M} \gamma(t)
\label{e3A-1},
\end{eqnarray}
where $\boldsymbol{P}$ is the impulsive force added at $t = 0$.
The analytical form of the relaxation function is obtained from a Stokes approximation (see the Appendix).

The simulations are performed with compressibility factors of $\varepsilon = 0.1$, 0.6, 1.0, and 1.5.
The condition $\varepsilon > 1$ implies that the sound propagation occurs more slowly than the viscous diffusion, according to the definition given in Eq.~(\ref{e1-1}).
The simulation results agree quite well with the analytical solutions for all of the compressibility factors $\varepsilon$, as shown in Fig.~\ref{f3}.
The oscillation of $t/\tau_\nu \gtrsim 2$ at $\varepsilon = 0.1$ arises from the periodic boundary conditions.
In this case, the sound pulse arrives at the end of the system at $t/\tau_\nu = 1.6$; 
afterward, the sound pulse returns and affects the particle motion.

The velocity relaxation results highlight some remarkable properties of dynamics in a compressible fluid.
In the case of an incompressible fluid, part of the particle momentum is instantly carried away by the propagation of the infinite-speed sound wave, 
and the particle moves as if its mass were $M^{\ast} = M + M_f/2$, where $M_f = 4\pi a^3\rho_0 /3$ is the mass of the displaced fluid~\cite{B12}.
Therefore, the velocity relaxation function for an incompressible fluid at the initial time is $\gamma(+0) = M/M^{\ast}$;
afterward, the particle velocity gradually decreases because of momentum diffusion in the fluid caused by its viscosity.
Eventually, this decay obeys the power law $t^{-3/2}$ in the long-time region.
On the other hand, in the case of a compressible fluid, the velocity relaxation function indicates that the relaxation due to sound propagation occurs in a finite time interval.
For the compressibility factor $\varepsilon < 1$, the two relaxation processes are almost separate, 
and the relaxation function coincides with that of an incompressible fluid after the relaxation due to the sound propagation occurs.
With an increase in the compressibility factor, nonmonotonic behavior is observed in the relaxation function, 
and finally, inversion of the particle velocity is observed, as shown in the investigation of the analytical solution~\cite{B7}.

The density deviation around the particle is shown in Fig.~\ref{f4}, for which an analytical form has not been obtained.
For a small compressibility $\varepsilon = 0.1$, the sound wave pulse expands from the particle very quickly, 
which corresponds to a rapid decrease in the particle momentum due to sound propagation.
On the other hand, for a large compressibility $\varepsilon = 1.5$, the sound pulse does not spread.
The pulse remains within the vicinity of the particle and gradually decays through viscous diffusion.
The continuation of high fluid density in front of the particle is expected to cause backtracking motion.

\begin{figure}[bp]
\includegraphics[width=42mm]{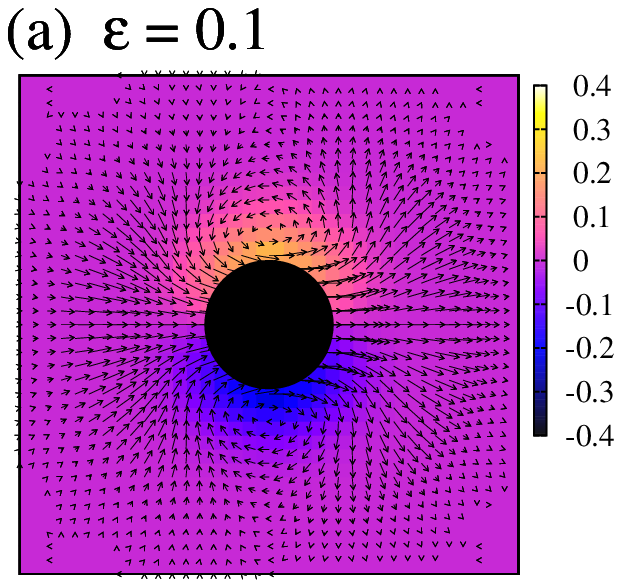}
\includegraphics[width=42mm]{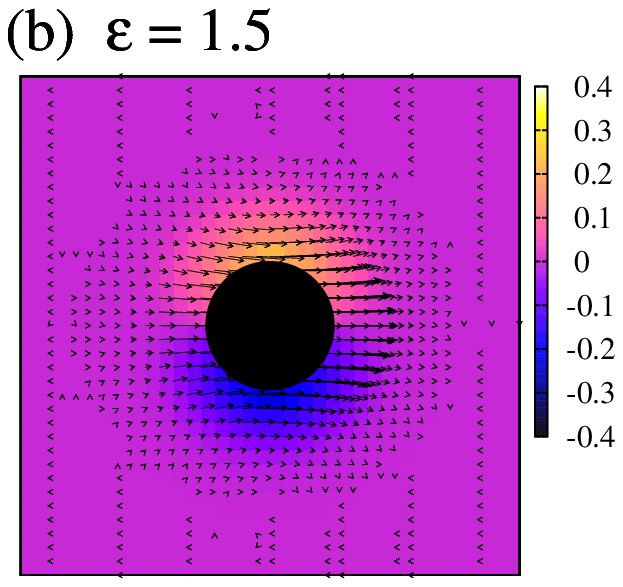}
\caption{\label{f5} (Color online) The fluid velocity field and its rotation, $\boldsymbol{\nabla} \times \boldsymbol{v}$, at $t/\tau_\nu = 0.27$.
The values are normalized by the factor $M a^{\ast}/|\boldsymbol{P}|$.
The component normal to the figure plane is depicted.
The compressibility factors are (a) $\varepsilon = 0.1$ and (b) $\varepsilon = 1.5$.
The color scale (gray scale) represents negative (darker) to positive (lighter) rotation.
The black circle represents the particle.
The direction of the initial particle velocity is right in these pictures.
}
\end{figure}

\begin{figure}[tbp]
\includegraphics[width=42mm]{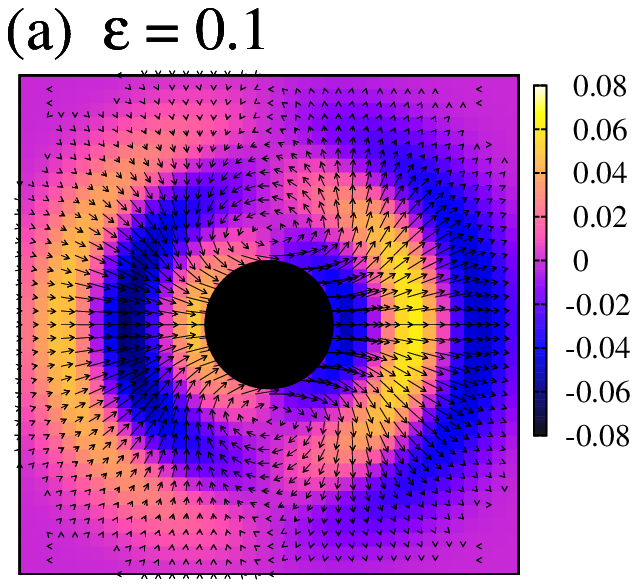}
\includegraphics[width=42mm]{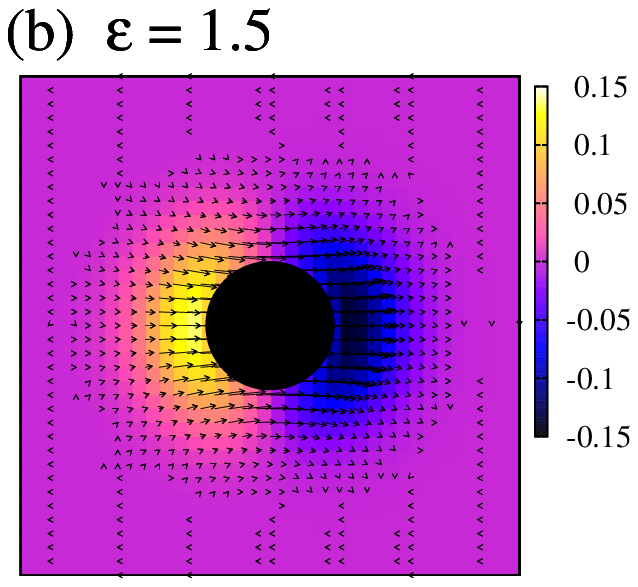}
\caption{\label{f6} (Color online) The fluid velocity field and its divergence, $\boldsymbol{\nabla} \cdot \boldsymbol{v}$, at $t/\tau_\nu = 0.27$.
The values are normalized by the factor $M a^{\ast}/|\boldsymbol{P}|$.
The compressibility factors are (a) $\varepsilon = 0.1$ and (b) $\varepsilon = 1.5$.
The color scale (gray scale) represents negative (darker) to positive (lighter) divergence.
The black circle represents the particle.
The direction of the initial particle velocity is right in these pictures.
}
\end{figure}

The velocity field around the particle is also affected by the compressibility.
The velocity field can be formally decomposed into two components:
an incompressible (solenoidal) component $\boldsymbol{v}^{\rm{I}}$ and a compressible component $\boldsymbol{v}^{\rm{C}}$.
The former is a solenoidal vector field when $\boldsymbol{\nabla} \cdot \boldsymbol{v}^{\rm{I}} = 0$, and the latter is an irrotational vector field when $\boldsymbol{\nabla} \times \boldsymbol{v}^{\rm{C}} = 0$.
Therefore, the divergence of the velocity field $\boldsymbol{\nabla} \cdot \boldsymbol{v}$ gives information regarding the compressible component, 
and the rotation of the velocity field $\boldsymbol{\nabla} \times \boldsymbol{v}$ gives information pertaining to the incompressible component.
The rotation and divergence of the velocity field are shown in Figs.~\ref{f5} and \ref{f6}, respectively.
The velocity field is also depicted in these figures.
An obvious vortex ring is observed around the particle for a small compressibility $\varepsilon = 0.1$,
while the vortex ring is not clear for a large compressibility $\varepsilon = 1.5$.
The rotation corresponds to the intensity of the vorticity, and the map in Fig.~\ref{f5} represents a pair of vortex rings that is inherent to an incompressible fluid~\cite{B13}.
The rotation remains nearly constant, regardless of the compressibility,
which indicates that the incompressible component is not affected by the compressibility.
The vector fields $\boldsymbol{v}^{\rm{I}}$ and $\boldsymbol{v}^{\rm{C}}$ influence each other only through the nonlinear terms in Eqs.~(\ref{e2A-3}) and (\ref{e2A-4});
however, in the present simulations, a low Reynolds number flow is assumed, and the additivity of the incompressible and compressible components is almost valid.
Therefore, the effect of compressibility is observed only from the divergence of the velocity field depicted in Fig.~\ref{f6}.
The region of positive divergence represents the source of the flow, and the negative region represents the sink of the flow.
From the equation of continuity Eq.~(\ref{e2A-3}), the divergence is equal to the reverse sign of the time change rate of the mass density:
\begin{eqnarray}
\boldsymbol{\nabla} \cdot \boldsymbol{v} &=& -\frac{D}{Dt}\ln \rho
\label{e3A-2}.
\end{eqnarray}
According to this relation, the source corresponds to the density decrease and the sink corresponds to the density increase.
Therefore, the source and the sink move according to the sound propagation.
The propagation speed decreases with increasing compressibility, 
which corresponds to the difference in Fig.~\ref{f6} related to the compressibility.
The pattern of the total velocity field is described as the superposition of the vortex convection of the incompressible component $\boldsymbol{v}^{\rm{I}}$
and the source-sink flow of the compressible component $\boldsymbol{v}^{\rm{C}}$.
The compressibility factor governs the relative time evolution of each component to produce various flow patterns.

\subsection{Fluctuation}

\begin{figure}[tbp]
\includegraphics[width=90mm]{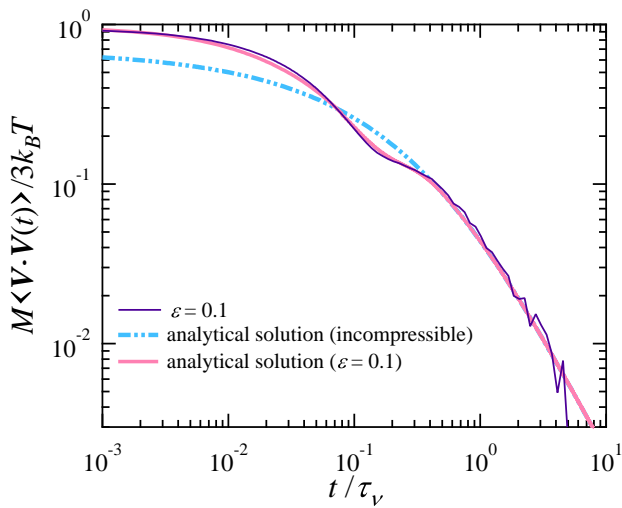}
\caption{\label{f7} (Color online) The velocity autocorrelation function at $\varepsilon = 0.1$ and $k_B T = 10^{-4}$.
The bold solid line represents the analytic solution of the velocity relaxation function.
The bold dashed double-dotted line shows the analytic solution for an incompressible fluid.
}
\end{figure}

Thermal fluctuations are introduced through random stress in the host fluid.
Computationally, a random stress term satisfying Eq.~(\ref{e2A-9}) is added to the stress tensor given by Eq.~(\ref{e2A-5}).
We consider a single particle in the fluid, which moves randomly as a result of thermal fluctuations.
From the fluctuation-dissipation theorem, the velocity autocorrelation function of the particle is related to the relaxation function as
\begin{eqnarray}
\gamma(t) &=& \frac{M}{3k_B T} \langle \boldsymbol{V}(0) \cdot \boldsymbol{V}(t) \rangle
\label{e3B-1}.
\end{eqnarray}
The accuracy of the fluctuating system can be confirmed by the validity of this relation.
In the numerical procedure for solving a stochastic differential equation, 
the numerical error can be larger than that in an ordinary differential equation 
due to the truncation error in the time integration of the random noise term~\cite{B14}.
This error is decreased with the decrease in the time increment, 
and we set the smaller time increment than that in the relaxation case as $h = 0.01$.
The system size is therefore scaled down to $L_x \times L_y \times L_z = 64 \times 64 \times 64$
to compensate the increased computational demand due to the small $h$
explained above.
The simulation results for $\varepsilon = 0.1$ are shown in Fig.~\ref{f7}, 
which shows good agreement with the analytical solution of the relaxation function.
The consistency between the input and calculated temperatures is also tested.
We can evaluate the temperature from the average kinetic energy of a fluid or a particle, 
i.e., $k_B T_f = \Delta^3 \langle \rho \boldsymbol{v}^2\rangle /3$ or $k_B T_p = M \langle \boldsymbol{V}^2\rangle /3$.
We evaluated these two temperatures as the ratio to the input temperature $T$.
From the fluid motion, the temperature was evaluated as $T_f/T = 1.06$, while that of the particle motion was $T_p/T = 0.93$.
The overestimation of the fluid temperature $T_f$ is simply due to the 
truncation error in the time integration of the stochastic differential equation.
On the other hand, the particle temperature $T_p$ is slightly below
the input temperature $T$.
This discrepancy is considered to be due to the small numerical inaccuracy 
introduced in treating 
the momentum transfer through the particle-fluid interface using the SPM.
Further improvements on the treatments of fluctuations in
the particle-fluid interfacial region will be discussed in the future.

\section{Conclusion}

We extended the SPM to particle dispersions in 
compressible fluids.
The validity of the method was confirmed by calculating the velocity relaxation function of a single spherical particle in a compressible fluid.
The effect of compressibility on the velocity relaxation was also observed, showing two-stage relaxation in a low-compressibility fluid and backtracking motion in a high-compressibility fluid.
These particle motions were considered by investigating the fluid density deviation.
The propagation of the sound pulse around the particle is governed by the compressibility, 
and the influence of the sound disappears in a low-compressibility fluid but is maintained in a high-compressibility fluid.
The effect of compressibility on the fluid velocity field was also observed, 
which was essentially understood to arise from changes in the time evolution of the source-sink flow component caused by the compressibility.

A simulation of the motion of a single spherical particle in a fluctuating fluid was also performed.
The calculated velocity autocorrelation function of the particle showed good agreement with the analytical solution of the relaxation function, 
and the validity of the fluctuation-dissipation theorem was confirmed.
%On the other hand, slight inconsistencies between the temperatures estimated from the particle motion and the fluid motion were observed.
%This discrepancy is considered to be due to the incomplete treatment of the particle-fluid interface.

\begin{acknowledgments}
The authors would like to express their gratitude to 
Dr. S. Yasuda, Dr. H. Kobayashi, and Dr. Y. Nakayama 
for their useful comments and discussions.
This work was supported by KAKENHI 23244087 and the JSPS Core-to-Core
Program ``International research network for non-equilibrium dynamics of
soft matter.''
\end{acknowledgments}

\section*{Appendix: Velocity Relaxation of a Particle in a Compressible Fluid}
\renewcommand{\theequation}{A\arabic{equation}}
\setcounter{equation}{0}

The equation of motion for a spherical particle under an external force $\boldsymbol{E}(t)$ in a compressible fluid can be expressed by the linear response theory as
\begin{eqnarray}
M \frac{\mathrm{d}}{\mathrm{d} t} \boldsymbol{V} &=& -\int^t_{-\infty} \mathrm{d} s \zeta(t-s) \boldsymbol{V}(s) + \boldsymbol{E}(t)
\label{eA-1},
\end{eqnarray}
where $\zeta(t)$ is the memory kernel of the friction force.
The memory kernel is analytically expressed in frequency representation.
Assuming a stick boundary condition on the surface of the sphere, the memory kernel is obtained from the linearized hydrodynamic equations (Stokes approximation) as~\cite{B5,B6}
\begin{eqnarray}
\hat{\zeta}(\omega) &=& \int_{-\infty}^{\infty} \zeta(t) e^{i\omega t} \mathrm{d} t,\\
&=& \frac{4\pi}{3} \eta ax^2 \nonumber \\
&&  \times \frac{(1+x)(9-9iy-2y^2) + x^2 (1-iy)}{2x^2 (1-iy) - (1+x)y^2 - x^2 y^2}
\label{eA-2},
\end{eqnarray}
with 
\begin{eqnarray}
x &=& a(-i\omega \rho_0 / \eta)^{1/2},\ \ \ \ y = a \omega / \tilde{c}
\label{eA-3},
\end{eqnarray}
and
\begin{eqnarray}
\tilde{c} &=& \left[ c^2 - \frac{i \omega}{\rho_0} \left( \frac{4}{3} \eta + \eta_v \right) \right]^{1/2}
\label{eA-4}.
\end{eqnarray}
According to Eq.~(\ref{eA-1}), the particle velocity is linearly dependent on the external force in frequency representation:
\begin{eqnarray}
\hat{\boldsymbol{V}}(\omega) &=& \hat{\Gamma}(\omega) \hat{\boldsymbol{E}}(\omega)
\label{eA-5},
\end{eqnarray}
where the admittance $\hat{\Gamma}(\omega)$ is given by
\begin{eqnarray}
\hat{\Gamma}(\omega) &=& [-i\omega M + \hat{\zeta}(\omega)]^{-1}
\label{eA-6}.
\end{eqnarray}
In the case that an impulsive force is exerted on the sphere, 
the external force in frequency representation is given as a constant vector $\hat{\boldsymbol{E}}(\omega) = \boldsymbol{P}$, 
and the velocity relaxation function is given by $\gamma(t) = M\Gamma(t)$ according to Eq.~(\ref{e3A-1}).

\nocite{*}

\bibliography{apssamp}% Produces the bibliography via BibTeX.

\begin{thebibliography}{9}
\bibitem{B1} Y. Nakayama and R. Yamamoto, Phys. Rev. E {\bf 71}, 036707 (2005).
\bibitem{B2} Y. Nakayama, K. Kim, and R. Yamamoto, Eur. Phys. J. E {\bf 26}, 361 (2008).
\bibitem{B3} R. Zwanzig and M. Bixon, Phys. Rev. A {\bf 2}, 2005 (1970).
\bibitem{B4} H. Metiu, D. W. Oxtoby, and K. F. Freed, Phys. Rev. A {\bf 15}, 361 (1977).
\bibitem{B5} D. Bedeaux and P. Mazur, Physica {\bf 78}, 505 (1974).
\bibitem{B6} B. U. Felderhof, Phys. Fluids {\bf 19}, 126101 (2007).
\bibitem{B7} B. U. Felderhof, J. Chem. Phys. {\bf 123}, 044902 (2005).
\bibitem{B8} L. D. Landau and E. M. Lifshitz, {\it Fluid Mechanics} (Pergamon, London, 1959).
\bibitem{B9} T. Iwashita, Y. Nakayama, and R. Yamamoto, J. Phys. Soc. Jpn. {\bf 77}, 074007 (2008).
\bibitem{B10} G. Erlebacher, M. Y. Hussaini, H. O. Kreiss, and S. Sarkar, Theor. Comput. Fluid. Dyn. {\bf 2}, 73 (1990).
\bibitem{B11} H. Hashimoto, J. Fluid Mech. {\bf 5}, 317 (1959).
\bibitem{B12} R. Zwanzig and M. Bixon, J. Fluid. Mech. {\bf 69}, 21 (1975).
\bibitem{B13} B. U. Felderhof, Phys. Fluids {\bf 19}, 073102 (2007).
\bibitem{B14} R. L. Honeycutt, Phys. Rev. A {\bf 45}, 600 (1992).
%\bibitem{B15} A. S. Dukhin and P. J. Goetz, {\it Ultrasound for Characterizing Colloids: Particle Sizing, Zeta Potential, Rheology}, (Amsterdam, Tokyo, Elsevier, 2002).

\end{thebibliography}

\end{document}